\documentclass[prd,showpacs,preprintnumbers,amsmath,amssymb]{revtex4}
\usepackage{graphicx}% Include figure files
\usepackage{dcolumn}% Align table columns on decimal point
\usepackage{bm}% bold math
\begin{document}

\title{QED Effective Action in Magnetic Field Backgrounds and Electromagnetic Duality}

\author{Sang Pyo Kim}\email{sangkim@kunsan.ac.kr}
\affiliation{Department of Physics, Kunsan National University, Kunsan 573-701, Korea}
\affiliation{Institute of Astrophysics, Center for Theoretical Physics,Department of Physics, National Taiwan University, Taipei 106, Taiwan}
\affiliation{Yukawa Institute for Theoretical Physics, Kyoto University, Kyoto 606-8502, Japan}

\medskip

\date{\today}

\begin{abstract}
In the in-out formalism we advance a method of the inverse scattering matrix for calculating effective actions in pure magnetic field backgrounds.
The one-loop effective actions are found in a localized magnetic field of Sauter type and approximately in
a general magnetic field by applying the uniform semiclassical approximation.
The effective actions exhibit the electromagnetic duality between a constant electric field
and a constant magnetic field and between $E(x) = E \, {\rm sech}^2 (x/L)$ and $B(x) = B \, {\rm sech}^2 (x/L)$.
\end{abstract}

\maketitle

\section{Introduction}\label{introduction}

The effective action in a background field probes the vacuum structure of the underlying theory.
In quantum electrodynamics (QED) the effective action in a constant electromagnetic field was first found by Heisenberg, Euler, Weisskopf \cite{heisenbergeuler} and later by Schwinger in the proper-time integral \cite{schwinger}.
The vacuum polarization by an electromagnetic field leads to prominent phenomena such as photon splitting,
direct photon-photon scattering, birefringence, and Schwinger pair production.
In spite of constant interest and continuous investigations, however, computing nonperturbative effective action beyond a constant
field has been regarded as a nontrivial, challenging task, and the exact effective actions have been known only for a few configurations of electromagnetic fields \cite{dittrichreuter}. The zeta-function regularization can be used for a constant magnetic field \cite{blviwi}, and the worldline integral \cite{rescsc}, the Green's function
(resolvent technique) \cite{chd,dunne-hall98-2,dunne-hall98}, and the lightcone coordinate \cite{fried} have also been introduced
to compute the effective actions in some electromagnetic fields.

In the in-out formalism based on the Schwinger variational principle \cite{schwinger51}, the vacuum persistence amplitude $
\langle {\rm out} \vert {\rm in} \rangle = e^{i \int d^3 x d t {\cal L}^{(1)} }$
leads to the effective action \cite{dewitt}
\begin{eqnarray}
\int d^3 x d t {\cal L}^{(1)} = \mp i \sum_{\bf K} \ln (\alpha_{\bf K}^*). \label{alp-act}
\end{eqnarray}
Here and throughout the paper the upper (lower) sign is for spinor (scalar) QED, $\alpha_{\rm K}$ is the
Bogoliubov coefficient between the in and out vacua, and
${\bf K}$ stands for all quantum numbers, such as momenta and/or energy and spin of the field.
The in-out formalism manifests the vacuum persistence relation
\begin{eqnarray}
2 {\rm Im} (\int d^3 x d t {\cal L}^{(1)}) &=& \mp  \sum_{\bf K} \ln (|\alpha_{\bf K}|^2) \nonumber\\
&=& \mp  \sum_{\bf K} \ln (1 \mp {\cal N}_{\bf K}), \label{vac per rel}
\end{eqnarray}
where ${\cal N}_{\bf K} = |\beta_{\bf K}|^2$ is the mean number of produced pairs and the Bogoliubov relation has been used:
\begin{eqnarray}
|\alpha_{\bf K}|^2 \pm |\beta_{\bf K}|^2 =1.
\end{eqnarray}
Thus the vacuum persistence relation relates the mean number of pairs to the imaginary part of the effective action.
It was shown in Ref. \cite{ahn} that the vacuum persistence amplitude could provide the QED effective action.
Using the gamma-function regularization, Kim, Lee, and Yoon have further developed the effective actions in the in-out formalism
in a temporally or a spatially localized electric field of Sauter-type \cite{kly08,kim09,kly10} and at finite temperature \cite{kly10-2}.
In the space-dependent gauge for electric fields the Bogoliubov coefficients from the second quantized field theory for
barrier tunneling \cite{nikishov70} are used to compute the effective actions \cite{kim09,kly10}.

The purpose of this paper is two-fold. First, we advance a method to find QED effective actions in static magnetic fields
in the in-out formalism. Second, we show the electromagnetic duality of QED actions between a constant electric field and a constant magnetic field
and also between a Sauter-type electric field and a magnetic field.
Contrary to a common belief that the in-in formalism should be used for pure magnetic fields,
we argue that the inverse scattering matrix for charged particles
may give the coefficient playing the same role as the Bogoliubov coefficient in the in-out formalism.
We use the inverse scattering matrix to find the effective actions in the proper-time integral
in a constant magnetic field and a Sauter-type magnetic field
$B(x) = B \, {\rm sech}^2 (x/L)$. Our effective action in the Sauter-type magnetic field is a multiple integral of proper time,
transverse momenta, and Euclidean energy while the effective action by
the resolvent Green function \cite{chd,dunne-hall98-2} involves a single integral of proper time.

The underlying idea is that the exponentially decreasing and increasing solutions in each asymptotic region for charged particles
in pure magnetic fields can be written in the form of Jost functions as for the scattering theory.
The normalizability condition on the Jost functions is the on-shell condition for physically bound states. In general, one set of
solutions in one asymptotic region can always be expressed through Jost functions by another set in the
other asymptotic region, which may be interpreted as the off-shell condition that extends scattering theory to bound states \cite{taylor}.
Remarkably the inverse scattering matrix, which is the ratio of the amplitude for exponentially increasing branch
to the amplitude for exponentially decreasing branch, plays the
analogous role for the Bogoliubov coefficient and leads to the effective actions in pure magnetic fields.
We illustrate this method for a constant magnetic field and a spatially localized magnetic field
of Sauter-type. We further show through one-loop effective actions in the in-out formalism that the electromagnetic duality holds between a constant electric field and a constant magnetic field and also between $E(x) = E \, {\rm sech}^2 (x/L)$ and $B(x) = B \, {\rm sech}^2 (x/L)$.
A common stratagem has been to show the duality of the Heisenberg-Euler and Schwinger effective action in the constant electromagnetic field under the dual transformation of electric field and magnetic field \cite{chopak}.

The organization of this paper is as follows.
In Sec. \ref{sec 4} we advance a method to find effective action from the inverse scattering matrix in a constant magnetic field
and compare it with the resolvent Green function method, and in Sec. \ref{sec 5} we apply the method to the Sauter-type magnetic field.
In Sec. \ref{sec 6} we show the duality of the one-loop effective actions in the constant and the Sauter-type electric and magnetic fields.
The Jost functions are derived for QED in magnetic fields in the Appendix.

\section{Constant Magnetic Field} \label{sec 4}

The transverse motion of a charged particle in a constant magnetic field
is confined to Landau levels and in a general configuration has both a discrete and a continuous spectrum of energy.
The vacuum defined by the lowest Landau level is stable and no pair is produced from pure magnetic fields due to
infinite instanton action and thereby the zero tunneling probability for the Dirac sea to decay \cite{kimpage02}.

The spin-diagonal Fourier component of the squared Dirac or Klein-Gordon equation \cite{schweber}
\begin{eqnarray}
\Bigl [\partial_x^2 - (k_y - q Bx)^2 + \omega^2 - m^2 - k_z^2 + 2\sigma (qB) \Bigr] \varphi_{(\sigma)} (x) = 0, \label{mag eq}
\end{eqnarray}
has the harmonic wave functions with the energy $\epsilon = \omega^2 - m^2 - k_z^2 + 2\sigma (qB)$ as bound states, corresponding
to the Landau levels $\epsilon = qB (2n+1)$. The general solutions for Eq. (\ref{mag eq}) are given by the parabolic cylinder functions
\begin{eqnarray}
D_p (\xi), \quad D_p(- \xi), \quad D_{-p-1} (i \xi), \quad D_{-p-1} (-i \xi), \label{con mag sol}
\end{eqnarray}
where
\begin{eqnarray}
\xi = \sqrt{\frac{2}{qB}} (k_y - qBx), \quad p = - \frac{1- 2\sigma}{2} + \frac{\omega^2 - m^2 - k_z^2}{2qB}.
\end{eqnarray}
In the Riemann sheet \cite{riemann}, the exponentially decreasing solutions are
$D_p (-\xi)$ at $x = \infty$ and $D_{p} (\xi)$ at $x = - \infty$ while the exponentially
increasing solutions are $D_{-p-1} (-i\xi)$ at $x = \infty$ and $D_{-p-1} (i\xi)$ at $x = - \infty$ (see the asymptotic formulas 9.246 of Ref. \cite{gr-table}).
As shown in the Appendix,
these functions can be used as the Jost functions for the bounded system as a generalization of scattering theory.
In fact, the connection formula 9.248 of Ref. \cite{gr-table} connects the bounded solution at $x = \infty$ with the solutions at $x = - \infty$
in terms of the Jost functions through Eq. (\ref{jos rel3}):
\begin{eqnarray}
D_{p} (- \xi) = \sqrt{2 \pi} \frac{e^{i(p+1) \frac{\pi}{2}}}{\Gamma (-p)} D_{-p-1} (i \xi) + e^{ip \pi} D_{p} (\xi).
\end{eqnarray}
We may introduce the inverse scattering matrix (\ref{in sc mat}), which is the ratio of the amplitude for the exponentially increasing
part to the amplitude for the exponentially decreasing part
\begin{eqnarray}
{\cal M}_p = \sqrt{2 \pi} \frac{e^{-i(p-1) \frac{\pi}{2}}}{\Gamma (-p)}. \label{sc mat}
\end{eqnarray}
Note that the scattering matrix in scattering theory \cite{taylor} is $1/{\cal M}_p$.

The inverse scattering matrix now carries the information about the potential and quantum states. For instance, the condition for bound states is
\begin{eqnarray}
{\cal M}_p = 0, \quad p = n, \quad (n = 0, 1, \cdots). \label{quan1}
\end{eqnarray}
The simple poles for the scattering matrix at physically bound states \cite{taylor} now become the simple zeros of $1/\Gamma (-p)$
for the inverse scattering matrix, for which $D_n(\xi)$ is the harmonic wave function up to a normalization constant with parity $e^{ip \pi}$.
That is, the nonnegative integer $p=n$ is the on-shell condition for Landau levels. Wick-rotating the time as $t = -i\tilde{t}$
and the frequency as $\omega = i \tilde{\omega}$, we observe that the inverse scattering matrix provides the effective action
in analogy with the in-out formalism for electric fields
\begin{eqnarray}
{\cal L}^{(1)} = \pm \frac{qB}{(2 \pi)} \sum_{\sigma} \int \frac{d \tilde{\omega}}{(2 \pi)} \frac{dk_z}{(2 \pi)} \ln ({\cal M}_p^*).
\end{eqnarray}
where the upper (lower) sign is for spinor (scalar) QED and $(qB)/(2 \pi)$ accounts for the wave packet centered around $k_y = qBx$.
Using the gamma-function regularization \cite{kly08,kim09,kly10,kim10},
summing over the spin states and carrying out the integration, we obtain the effective action of the standard result
\begin{eqnarray}
{\cal L}^{(1)} =  \mp \frac{1+ 2 |\sigma|}{2} \frac{(qB)^2}{(2\pi)^2}\int_{0}^{\infty} \frac{ds}{s^2} e^{- \frac{m^2s}{2qB}} F_{\sigma} (s),
\label{conB-eff}
\end{eqnarray}
where the spectral function is
\begin{eqnarray}
F_{\sigma} (s) = \frac{[\cosh(\frac{s}{2})]^{2 |\sigma|}}{\sinh(\frac{s}{2})} - \frac{2}{s} + (-1)^{2 |\sigma|} \frac{(1+ 2 |\sigma|)s}{12}.
\label{sp fun}
\end{eqnarray}
Here the Schwinger prescription has been employed for renormalization of the vacuum energy and the charge, which subtracts the divergent terms, the last two terms in Eq. (\ref{sp fun}), in the proper-time integral \cite{schwinger}. A passing remark is that the inverse scattering matrix is real and, therefore, the effective action does neither have an imaginary part
nor lead to the vacuum decay due to pair production.

We now compare the inverse scattering matrix method with the resolvent Green function method applied
to magnetic fields \cite{chd,dunne-hall98-2,dunne-hall98}, in which the effective action is
the sum and the trace of the resolvent Green function for Eq. (\ref{mag eq}).
Following Refs. \cite{chd,dunne-hall98-2} and choosing two independent solutions $D_p (\xi)$ and $D_p (- \xi)$, the effective action is given by
\begin{eqnarray}
{\cal L}^{(1)} &=& \pm \frac{qB}{(2 \pi)} \sum_{\sigma}
\int_{- \infty}^{\infty} \frac{d \tilde{\omega}}{(2 \pi)} \frac{dk_z}{(2 \pi)}
\frac{2\tilde{\omega}^2}{{\rm Wr}_x [D_p (\xi),D_p (-\xi)]} \int_{- \infty}^{\infty} dx D_p (\xi)D_p (-\xi) \nonumber\\
&=& \pm \frac{1}{2 (2 \pi)^3} \sum_{\sigma} \int_{- \infty}^{\infty} \tilde{\omega}^2 d \tilde{\omega}
dk_z \Bigl[\psi( \frac{1}{2} - \frac{p}{2}) + \psi (- \frac{p}{2})\Bigr]. \label{res-act}
\end{eqnarray}
Here we have used the formulas 7.711-2 and 8.335-1 of Ref. \cite{gr-table} in the second line and $\psi$ is the psi function $\psi(z) = d (\ln \Gamma(z))/dz$. Using the formula 8.361-1 of Ref. \cite{gr-table}
\begin{eqnarray}
\psi (z) = - \int_{0}^{\infty} ds \Bigl(\frac{e^{-zs}}{1- e^{-s}} - \frac{e^{-s}}{s} \Bigr),
\end{eqnarray}
summing over the spin states and performing the double Gaussian integral, we recover the standard effective action (\ref{conB-eff}) after the Schwinger prescription is done for renormalization. In the in-out formalism DeWitt has shown the equivalence of the effective action (\ref{alp-act})
from the Bogoliubov coefficient and the effective action from the Green function \cite{dewitt}.

\section{Spatially Localized Magnetic Field} \label{sec 5}

We now consider a spatially localized field $B(x) = B \, {\rm sech}^2(x/L)$ of Sauter-type along the $z$ direction with the space-dependent gauge field
\begin{eqnarray}
A_{\mu} = (0, 0, - BL \tanh (\frac{x}{L}), 0).
\end{eqnarray}
Then the spin-diagonal Fourier component of the squared Dirac or Klein-Gordon equation becomes \cite{schweber}
\begin{eqnarray}
\Bigl [\partial_x^2 - (k_y - q BL \tanh (\frac{x}{L}))^2 + \omega^2 - m^2 - k_z^2 +2 \sigma qB {\rm sech}^2 (x/L) \Bigr] \varphi_{(\sigma)} (x) = 0. \label{mag eq2}
\end{eqnarray}
The motion (\ref{mag eq2}) is bounded at $x = \pm \infty$, so the momentum $P_1$ along the longitudinal direction
takes an imaginary value, $P_{1(\pm)} = i\Pi_{1(\pm)}$,
\begin{eqnarray}
\Pi_{1(\pm)} (B) = \sqrt{(k_y \mp q BL)^2 - (\omega^2 - m^2 - k_z^2)}.
\end{eqnarray}
The solution may be found in terms of the hypergeometric function as
\begin{eqnarray}
\varphi_{(\sigma)} (x) = \xi^{\frac{L}{2} \Pi_{1(+)}} (1 - \xi)^{\frac{1-2\sigma}{2} + \lambda_{\sigma}} F(a, b; c; \xi), \label{saut sol}
\end{eqnarray}
where
\begin{eqnarray}
\xi = - e^{- 2 \frac{x}{L}}, \quad \lambda_{\sigma} = (qBL^2) \sqrt{1 + \Bigl(\frac{1-2|\sigma|}{2 qBL^2} \Bigr)^2},
\end{eqnarray}
and
\begin{eqnarray}
a &=& \frac{1-2\sigma}{2} + \frac{1}{2} (L\Pi_{1(+)}+ L\Pi_{1(-)} +2\lambda_{\sigma}) := \frac{1-2\sigma}{2} + \frac{\Omega_{(+)}}{2}, \nonumber\\
b &=& \frac{1-2\sigma}{2} + \frac{1}{2} (L\Pi_{1(+)} - L\Pi_{1(-)} +2\lambda_{\sigma}) := \frac{1-2\sigma}{2} + \frac{\Delta_{(+)}}{2}, \nonumber\\
c &=& 1 + L \Pi_{1(+)}.
\end{eqnarray}
The solution is bounded at $x = \infty \, (\xi = 0)$ since $\Pi_{1(+)}$ is positive. In the opposite limit $x = - \infty \, (\xi = - \infty)$,
using the connection formula 9.132 of Ref. \cite{gr-table}, we find the asymptotic form for the solution
\begin{eqnarray}
\varphi_{(\sigma)} = (-1)^{\frac{L}{2} \Pi_{1(+)}} \Bigl[
(- \xi)^{-\frac{L}{2} \Pi_{1(-)}} \frac{\Gamma (c) \Gamma(b -a)}{\Gamma(b) \Gamma(c-a)} + (- \xi)^{\frac{L}{2} \Pi_{1(+)}} \frac{\Gamma (c) \Gamma(a -b)}{\Gamma(a) \Gamma(c-b)} \Bigr]. \label{mag sa con}
\end{eqnarray}
The first term exponentially decreases while the second term increases. As $a > b > 0$ and $c > b$, the condition for bound states is
that $\Gamma (c-b)$ should be singular, which leads to a finite number of discrete spectrum
\begin{eqnarray}
c - b = \frac{L}{2} (\Pi_{1(+)} + \Pi_{1(-)}) - \lambda_{\sigma} + \frac{1+ 2 \sigma}{2} = - n, \quad (n =0, 1, \cdots), \label{quan2}
\end{eqnarray}
with $n + (1+ 2\sigma)/2 <\lambda_{\sigma}$.

Now we apply the method of the inverse scattering matrix in Sec. \ref{sec 4} and the Appendix. The asymptotic solutions (\ref{jos fn3}) are
$\xi^{L\Pi_{1(+)}/2}$ at $x = \infty$ and $(-\xi)^{-L\Pi_{1(-)}/2}$ at $x = - \infty$, so Eq. (\ref{mag sa con}) connects
the solutions in terms of the Jost functions (\ref{jos fn3}). Then the inverse scattering matrix is
\begin{eqnarray}
 {\cal M} = \frac{\Gamma (b) \Gamma(c-a)}{\Gamma(a) \Gamma (c-b)}, \label{sc-Bsa}
\end{eqnarray}
where $\Gamma (a-b)/\Gamma(b-a)$ that depend only $\Pi_{1(-)}$ can be gauged away by choosing $A_2 = BL (\tanh (x/L) + 1)$
and will not be included hereafter. In the limit of $qBL \gg |\omega|$, the inverse scattering matrix (\ref{sc-Bsa}) reduces to
$1/\Gamma(-p)$ with $p$ from Eq. (\ref{quan1}), modulo a term that is independent of the number of states and to be regulated away
through renormalization of the effective action. Applying the identity 8.334 of Ref. \cite{gr-table}, $
\Gamma (1-x) \Gamma (x) = \pi / \sin(\pi x)$, to negative values of
\begin{eqnarray}
c-a &=& \frac{1+ 2 \sigma}{2} + \frac{1}{2} (L\Pi_{1(+)} - L\Pi_{1(-)} - 2\lambda_{\sigma}) := \frac{1+2\sigma}{2} + \frac{\Delta_{(-)}}{2}, \nonumber\\
c-b &=& \frac{1+2\sigma}{2} + \frac{1}{2} (L\Pi_{1(+)} + L\Pi_{1(-)} - 2\lambda_{\sigma}) := \frac{1+ 2\sigma}{2} + \frac{\Omega_{(-)}}{2},
\end{eqnarray}
we may write the inverse scattering matrix as
\begin{eqnarray}
 {\cal M} = \frac{\Gamma (\frac{1-2\sigma}{2} + \frac{\Delta_{(+)}}{2}) \Gamma(\frac{1-2\sigma}{2} - \frac{\Omega_{(-)}}{2})}{\Gamma(\frac{1-2\sigma}{2} - \frac{\Delta_{(-)}}{2}) \Gamma (\frac{1-2\sigma}{2} + \frac{\Omega_{(+)}}{2})}. \label{sc-Bsa2}
\end{eqnarray}
Here we have deleted again the overall factor that is to be regulated away through the renormalization procedure.
Finally, the gamma-function regularization leads to the renormalized effective actions in the proper-time integral
\begin{eqnarray}
{\cal L}^{(1)} = \mp \frac{1+ 2 |\sigma|}{2}\int \frac{d \tilde{\omega}}{(2\pi)} \frac{d^2 {\bf k}_{\perp}}{(2 \pi)^2}
\int_{0}^{\infty} \frac{ds}{s}  \Bigl(e^{- \frac{\Omega_{(+)}}{2}s} + e^{\frac{\Delta_{(-)}}{2}s} - e^{\frac{\Omega_{(-)}}{2}s} - e^{- \frac{\Delta_{(+)}}{2}s} \Bigr) F_{\sigma} (s), \label{sa ma act}
\end{eqnarray}
with the same spectral function (\ref{sp fun}).
Here $\Omega_{(-)} <0 $ and $\Delta_{(-)} < 0$ and we have taken the Wick-rotation $t = - i \tilde{t}$ and $\omega = i \tilde{\omega}$ and used the Schwinger prescription for renormalization.
It would be interesting to compare and to show the equivalence of Eq. (\ref{sa ma act}) with Eq. (18) of Ref. \cite{dunne-hall98-2} from the resolvent Green function that has a single integral of proper time without the multiple integral of momenta and energy.
Following Ref. \cite{dunne-hall98}, the resolvent Green function from the solution (\ref{saut sol}) and another independent solution would result
in the effective action (\ref{sa ma act}) without the second and the last delta exponentials.

We briefly explain a scheme to find approximately the effective action for a general configuration of magnetic field,
when solutions are not known. The effective action may be used for measuring the inhomogeneity effect on birefringence in
a strong magnetic field. We assume a spatially localized field $B(x)$ along the $z$ direction and the gauge field $A_2$
such that $B(x) = \partial_x A_2(x)$.
Then the spin-diagonal Fourier component of the squared Dirac or Klein-Gordon equation will become
\begin{eqnarray}
\Bigl[ \partial_x^2  - \Pi_{1}^2 (x) \Bigr] \varphi_{(\sigma)} (x) = 0. \label{mag eq3}
\end{eqnarray}
where
\begin{eqnarray}
\Pi_{1}^2 (x) =  (k_y - q A_2(x))^2 - \omega^2 + m^2 + k_z^2 - 2 \sigma qB (x).
\end{eqnarray}
The uniform semiclassical approximation for electric fields \cite{dunne-hall98,kly10} suggests transforming (\ref{mag eq3}) into the form
\begin{eqnarray}
\Bigl[\partial_{\xi}^2 - \xi^2 + \frac{{\cal S}_{(\sigma)}}{\pi} + \frac{1}{(\partial_{x} \xi)^{3/2}} \partial_x^2 (\frac{1}{\sqrt{\partial_x \xi}}) \Bigr] w_{(\sigma)} (\xi) = 0, \label{b app eq}
\end{eqnarray}
where
\begin{eqnarray}
w_{(\sigma)} (\xi) = \sqrt{\partial_x \xi} \varphi_{(\sigma)} (x), \quad \Bigl(\xi^2 - \frac{{\cal S}_{(\sigma)}}{\pi} \Bigr) (\partial_x \xi)^2 = \Pi_{1}^2. \label{b ins}
\end{eqnarray}
The charged particle undergoes a periodic motion in the region $\Pi_{1}^2 \leq 0$, so the integration of Eq. (\ref{b ins}) over one period determines
the action \begin{eqnarray}
{\cal S}_{(\sigma)} = \oint \sqrt{- \Pi_{1}^2 (x)}dx.
\end{eqnarray}
Thus, in the approximation of neglecting the last term, Eq. (\ref{b app eq}) has the same form as Eq. (\ref{mag eq}) for the constant
magnetic field and approximately has the solutions $D_p (\sqrt{2} \xi)$, $D_p(- \sqrt{2}\xi)$, $D_{-p-1} (i \sqrt{2} \xi)$,
and $D_{-p-1} (-i \sqrt{2}\xi)$ with
\begin{eqnarray}
p = - \frac{1}{2} + \frac{{\cal S}_{(\sigma)}}{\pi}. \label{b app p}
\end{eqnarray}
Then the inverse scattering matrix is given by Eq. (\ref{sc mat}) with $p$ in Eq. (\ref{b app p}).
As $B(x) \leftrightarrow - B(x)$ under $k_y \leftrightarrow - k_y$ and $\sigma \leftrightarrow - \sigma$, we find the
unrenormalized effective action in a symmetric form
\begin{eqnarray}
{\cal L}^{(1)} &=& \mp \frac{1}{2} \sum_{\sigma} \int \frac{d \tilde{\omega}}{(2\pi)} \frac{d^2 {\bf k}_{\perp}}{(2 \pi)^2} \Bigl[ \ln \Gamma (- p (B) ) + \ln \Gamma (- p (-B) ) \Bigr]. \label{app B-eff}
\end{eqnarray}

We comment on how to estimate or improve the error in the approximate effective action (\ref{app B-eff}). Treating the last term of Eq. (\ref{b app eq}) as a perturbation, the exact solution is the sum of the homogeneous part, an approximate solution, and the inhomogeneous part, which is an integral equation of two independent approximate solutions and the perturbation. In the in-out formalism the effective action, which is the sum of
logarithm of the Bogoliubov coefficient or the inverse scattering matrix for each momentum and spin, becomes a functional of the exact solution to Eq. (\ref{mag eq3}).
Thus this systematic improvement combined with an appropriate renormalization scheme allows us to estimate the error in Eq. (\ref{app B-eff}) and also holds true for the temporally or spatially localized electric fields in Refs. \cite{kly08,kly10}, which is beyond the scope of this paper.

\section{Electromagnetic Duality of QED Actions} \label{sec 6}

To show the electromagnetic duality, we recapitulate the main results of Refs. \cite{kim09,kly10} for the constant electric
field and the Sauter-type electric field $E(z) = E \, {\rm sech}^2(z/L)$ along the $z$ direction in the space-dependent gauge
\begin{eqnarray}
A_{\mu} = (A_0(z), 0, 0, 0).
\end{eqnarray}
First, in the constant electric field with $A_0 (z) = - Ez$, the Bogoliubov coefficient in the in-out formalism is given by
\begin{eqnarray}
\alpha_{(r)} = \sqrt{2 \pi} \frac{e^{-i(2p^*+p+1) \frac{\pi}{2}}}{\Gamma (-p)}, \quad p = - \frac{1 + r}{2} + i \frac{m^2 + {\bf k}_{\perp}^2}{2qE}.
\end{eqnarray}
Here $r$ is the eigenvalues for $\sigma^{03}= (i/4) [\gamma^{0}, \gamma^{3}]$. Thus, we find
the effective action
\begin{eqnarray}
{\cal L}^{(1)} &=&  \pm i \frac{qE}{2(2 \pi)} \sum_{\sigma r} \int \frac{d^2 {\bf k}_{\perp}^2}{(2\pi)^2}
\int_{0}^{\infty} \frac{ds}{s} \frac{e^{-p^*s}}{1 - e^{-s}} \nonumber\\
&=& \mp \frac{1+ 2|\sigma|}{2} \frac{(qE)^2}{(2 \pi)^2} \int_{0}^{\infty} \frac{ds}{s^2} e^{- \frac{m^2 s}{2qE}} (i F_{\sigma} (is)).
\end{eqnarray}
In the second line the Schwinger prescription has been used to obtain the renormalized effective action.
The $\Gamma (-p^*)$ in $\alpha_{(r)}^*$ for the effective action is the same as that from the inverse scattering matrix (\ref{sc mat}),
provided that $E = i B$ and $\omega = i \tilde{\omega}$. This implies that the unrenormalized and renormalized
effective actions for the electric field are dual to those for the magnetic field. Note that $E = i B$ is consistent with duality of
the convergent series of the Heisenberg-Euler and Schwinger effective action \cite{chopak} and that the electromagnetic duality
can also be shown in the resolvent Green function method by comparing Eq. (\ref{res-act}) with Eq. (22) of Ref. \cite{dunne-hall98}.

Second, for the Sauter-type electric field the space-dependent gauge field is $A_{0} = - EL \, \tanh (z/L)$.
The effective action in the electric field $E(t) = E \, {\rm sech}^2(t/T)$ was studied in Refs. \cite{kly08,dunne-hall98}
and in the electric field $E(z) = E \, {\rm sech}^2(z/L)$ in Ref. \cite{kly10}. The spin-diagonal Fourier component of the squared
Dirac or Klein-Gordon equation has the asymptotic longitudinal momentum along the $z$ direction
\begin{eqnarray}
P_{3(\pm)} (E) = \sqrt{(\omega \mp qEL)^2 - (m^2+ k_x^2 + k_y^2)}.
\end{eqnarray}
Then, the Bogoliubov coefficient $\alpha_{(r)}^*$ in Eq. (20) of Ref. \cite{kly10}, under
$E = iB$ and $\omega=i \tilde{\omega}$ and under the interchange of $\tilde{\omega} \leftrightarrow -k_y$
and $k_x \leftrightarrow k_z$ and in the Riemann sheet \cite{riemann}, has the same arguments for the gamma functions in Eq. (\ref{sc-Bsa})
\begin{eqnarray}
P_{3 (\pm)} (E) = - i \Pi_{1(\pm)} (B), \quad \lambda_{r} (E) = -i \lambda_{\sigma} (B)
\end{eqnarray}
and thus
\begin{eqnarray}
\Omega_{(\pm)} (E) = - i \Omega_{(\pm)} (B), \quad \Delta_{(\pm)} (E) = - i \Delta_{(\pm)} (B).
\end{eqnarray}
The fact that the Bogoliubov coefficient in the Sauter-type electric field has the same form as the inverse scattering matrix (\ref{sc-Bsa})
in the Sauter-type magnetic field shows that the unrenormalized and renormalized effective actions can be analytically continued
from one field to the other and are dual to each other under $E = iB$.

\section{Conclusion} \label{conclusion}

In this paper we have proposed a method for finding the one-loop effective action in magnetic field backgrounds in the in-out formalism, in which
the effective action is the logarithm of the Bogoliubov coefficient in the second quantized field theory.
The in-out formalism may have a further extension to magnetic fields when the inverse scattering matrix
is used for the Bogoliubov coefficient. As shown in Secs. \ref{sec 4}, \ref{sec 5} and the Appendix, the Jost functions
for off-shell solutions give the inverse scattering matrix, which is the ratio of the amplitude of the exponentially increasing
part to the amplitude of the exponentially decreasing part and plays an analogous role of the Bogoliubov coefficient for
electric fields in the space-dependent gauge. In fact, in the space-dependent gauge the Bogoliubov coefficient
is determined by the Jost functions for tunneling solutions
through barriers in electric fields while the inverse scattering matrix is determined by the Jost functions
for off-shell solutions in magnetic fields, both of which are the Wronskian for the solutions with required behaviors in two asymptotic regions.

We have illustrated the method by computing the effective actions in a constant magnetic
field and a localized magnetic field of Sauter type. As the Bogoliubov coefficient for the electric field
and the inverse scattering matrix for the magnetic field are determined by
Jost functions, they can analytically continue to each other under the electromagnetic duality,
and the QED effective actions, renormalized or unrenormalized, exhibit duality between electric and magnetic fields of the same profile.
We have explicitly showed the electromagnetic duality of QED effective actions in constant and Sauter-type electric
and magnetic fields. As QED effective actions in time-varying or spatially localized fields are nontrivial,
the new method in the in-out formalism may provide an alternative scheme to understanding the vacuum structure.

\acknowledgments

The author would like to thank Don N. Page, Hyun Kyu Lee, and Yongsung Yoon for early collaborations,
and W-Y. Pauchy Hwang and Wei-Tou Ni for helpful discussions.
He also thanks Misao Sasaki for the warm hospitality at Yukawa Institute for Theoretical Physics, Kyoto University
and Christian Schubert at Instituto de F\'{i}sica y Matem\'{a}ticas, Universidad Michoacana
de San Nicol\'{a}s de Hidalgo. This work was supported in part by Basic Science Research Program through
the National Research Foundation (NRF) funded by the Korea Ministry of Education, Science and Technology (2011-0002-520)
and in part by National Science Council Grant (NSC 100-2811-M-002-012) by the Taiwan government.

\appendix

\section{Jost Functions for QED in Magnetic Field} \label{sec mag-ga}

The spin-diagonal Fourier component of the squared Dirac or Klein-Gordon equation in a magnetic field
becomes the one-dimensional Schr\"{o}dinger equation for a bounded system
\begin{eqnarray}
[\partial_{z}^2 - {\kappa}^2 + V (z) ] \varphi_{\kappa} (z) = 0, \label{B-jos3}
\end{eqnarray}
where $V \geq 0$ and $- {\kappa}^2 + V(\pm \infty) = - {\kappa}^2_{(\pm)}$.
We may introduce the exponentially decreasing solutions in each asymptotic region
\begin{eqnarray}
f(z, \kappa) \stackrel{z = \infty}{\longrightarrow} \frac{e^{-\kappa_{(+)} z}}{\sqrt{2 \kappa_{(+)}}}, \quad
g(z, \kappa) \stackrel{z = - \infty}{\longrightarrow} \frac{e^{\kappa_{(-)} z}}{\sqrt{2 \kappa_{(-)}}}, \label{jos fn3}
\end{eqnarray}
and the exponentially increasing solutions $f(z,-\kappa)$ and $g(z,-\kappa)$ in their region as independent solutions. Each set of solutions
satisfies the Wronskian
\begin{eqnarray}
[f(z,\kappa), f(z, -\kappa)] = 1, \quad [g(z,\kappa), g(z, -\kappa)] = -1. \label{jos wr3}
\end{eqnarray}
We may then express each set of solutions in terms of the other set as
\begin{eqnarray}
f(z, \kappa) = C_1 (\kappa) g(z, \kappa) + C_2 (\kappa) g(z, - \kappa), \nonumber\\
g(z, \kappa) = \tilde{C}_1 (\kappa) f(z, \kappa) + \tilde{C}_2 (\kappa) f(z, -\kappa), \label{jos rel3}
\end{eqnarray}
where the Jost functions are determined from (\ref{jos wr3})
\begin{eqnarray}
C_1 (\kappa) &=& - [f(z,\kappa), g(z,-\kappa)] = - \tilde{C}_1 (-\kappa), \nonumber\\
C_2 (\kappa) &=&  [f(z,\kappa), g(z,\kappa)] = \tilde{C}_2 (-\kappa). \label{bog coef3}
\end{eqnarray}
It follows that
\begin{eqnarray}
C_2 (\kappa) C_2 (- \kappa) - C_1 (\kappa) C_1 (- \kappa) = 1. \label{jos bog3}
\end{eqnarray}
Finally, we define the inverse scattering matrix as the ratio of the amplitude for the exponentially increasing part to the amplitude
for the exponentially decreasing part, which is the inverse of the scattering matrix \cite{taylor} and now is given by
\begin{eqnarray}
{\cal M}_{\kappa} = \frac{C_2 (\kappa)}{C_1(\kappa)}. \label{in sc mat}
\end{eqnarray}

\end{document}